\documentclass[3p,sort&compress]{elsarticle}

\usepackage{graphicx}
\usepackage{amsmath}
\usepackage{amssymb}
\usepackage{mathrsfs}
\usepackage{subfigure}
\def\be{\begin{equation}}
\def\ee{\end{equation}}
\def\ba{\begin{array}}
\def\ea{\end{array}}
\def\bea{\begin{eqnarray}}
\def\eea{\end{eqnarray}}
\def\no{\nonumber}
\def\({\left(}
\def\){\right)}
\def\[{\left[}
\def\]{\right]}

\usepackage{lineno}

\journal{Journal of \LaTeX\ Templates}

\begin{document}
\begin{frontmatter}

\title{Generation and controlling of ultrashort self-similar solitons and rogue waves in inhomogeneous optical waveguide}

\author[mymainaddress,mysecondaryaddress]{Harneet Kaur}

\author[mymainaddress,mysecondaryaddress2]{Nisha}

\author[mysecondaryaddress2]{Amit Goyal\corref{mycorrespondingauthor}}
\cortext[mycorrespondingauthor]{Corresponding author}
\ead{amit.goyal@ggdsd.ac.in}

\author[mysecondaryaddress3]{Thokala Soloman Raju}

\author[mymainaddress]{C. N. Kumar}

\address[mymainaddress]{Department of Physics, Panjab University, Chandigarh 160014, India}
\address[mysecondaryaddress]{Department of Physics, Government College for Women, Karnal 132001, India}
\address[mysecondaryaddress2]{Department of Physics, GGDSD College, Chandigarh 160030, India}
\address[mysecondaryaddress3]{Indian Institute of Science Education and Research (IISER) Tirupati, Andhra Pradesh 517507, India}

\begin{abstract}
We present exact bright, dark and rogue soliton solutions of generalized higher-order nonlinear Schr\"{o}dinger equation, describing the ultrashort beam propagation in tapered waveguide amplifier, via a similarity transformation connected with the constant-coefficient Sasa-Satsuma and Hirota equations. Our exact analysis takes recourse to identify the allowed tapering profile in conjunction with appropriate gain function which corresponds to $\mathcal{PT}$-symmetric waveguide. We extend our analysis to study the effect of tapering profiles and higher-order terms on the evolution of self-similar waves and thus enabling one to control the self-similar wave structure and dynamical behavior.
\end{abstract}

\end{frontmatter}


\section{Introduction}
Self-similar waves have attracted the attention of researchers in the field of nonlinear optics mainly due to their applications in high power pulse amplification, photonics, optical communication and optical signal processing \cite{krug1,pono2}. Optical self-similar waves appears as a result of delicate balance among model parameters such as nonlinearity, dispersion, gain and inhomogeneity. They maintain their overall shape but their width, amplitude and phase chirp changes with model parameters \cite{krug2,pono1,antar}. Due to these properties, self-similar solitons, also known as similaritons, are studied extensively in Yb doped amplifier \cite{krug1}, laser resonator \cite{ild}, graded-index waveguide \cite{pono1,wu} and coupled systems \cite{hli}. Over the last few decades, there is a phenomenal growth in the study of similaritons \cite{goyal1,goyal2,fzhang,choudhuri,he} and self-similar rogue waves \cite{dai2010,pra1,pra2} in nonlinear optics. In literature, optical self-similar waves have been studied in various forms such as Hermite Gaussian, compact parabolic and hybrid functions \cite{krug1,chang}, exact solutions \cite{krug2,pono1}, and quasisoliton solutions \cite{wu}.
\par Over the past years, the concept of self-similarity has propelled the development of ultrafast nonlinear optics \cite{dud}, specifically in the field of ultrashort self-similar pulse propagation \cite{ultrashort1,ultrashort2} and ultrafast fiber lasers based on self-similar pulse evolution \cite{ultrashort3}. The ultrashort optical pulses can be realized  by incorporating the higher-order effects such as third-order dispersion, self-steepening, self-frequency shift to the standard nonlinear Schr\"{o}dinger equation known as higher order nonlinear Schr\"{o}dinger equation (HNLSE) \cite{kodama1}. The constant coefficient HNLSE has been shown to support bright / dark soliton \cite{bd2,bd3} and rogue wave solutions \cite{rw3} under different parametric conditions. Recently, dipole solitons \cite{dipole} and periodic soliton interactions \cite{liu1,liu2} has been studied in the presence of higher-order effects. A significant work has also been done to obtain soliton-like and rogue wave solutions for variable-coefficient HNLSE with inhomogeneous gain / loss media \cite{hao,dai2012}. Subsequently, authors have studied the soliton propagation in inhomogeneous fiber system modeled by generalized variable-coefficient HNLSE incorporating the terms correspond to complex potential \cite{hp2,hp3}. The inhomogeneous complex optical waveguides, modeled by complex potential with special mathematical property that real part is an even function and imaginary part is an odd function of position, are known as parity-time ($\mathcal{PT}$) symmetric waveguides. In recent years, a significant research has been done on the evolution of solitons for wave propagation in $\mathcal{PT}$-symmetric optical waveguides modeled by nonlinear Schr\"{o}dinger equation with complex potential \cite{mussli2008,shi,Khare} and in the presence of one or more higher-order effects \cite{chen,dai4,saha}. The notion of  $\mathcal{PT}$-symmetry was originated from quantum mechanics \cite{bender} and implemented in nonlinear optics by Ganainy et al. \cite{gana}. Most of the work has been done by considering the optical waveguide to be $\mathcal{PT}$-symmetric along transverse direction \cite{mussli2008,shi,Khare,chen,dai4,saha,ruter,wim}, but recently authors have considered the $\mathcal{PT}$-symmetry property along longitudinal direction in different contexts \cite{nix,wang2,shailza}. Nixon et al. \cite{nix} studied the light propagation in optical waveguide by considering the $\mathcal{PT}$-symmetry property along longitudinal direction. Motivated by these works, we consider the beam propagation in optical waveguide satisfying $\mathcal{PT}$-symmetry property in longitudinal direction, such that real part of the complex potential describes the geometry of tapered waveguide and imaginary part represents the gain / loss media.

\par The properties and the impact of tapered profiles in optical waveguides have been studied extensively, both theoretically and experimentally \cite{supercontinuum,coupling,raman,tap6,tap5,tap7,tap1,tap4,tap2,tap3}. Tapering finds applications in extended broadband supercontinuum generation \cite{supercontinuum}, reduction of the reflection losses and mode mismatch \cite{coupling} and Raman amplification \cite{raman}. Recently, there is a considerable interest on self-similar waves in tapered graded-index waveguides \cite{pono1,wu,tap4}, compression and amplification of ultrashort pulses in tapered waveguides \cite{tap6,tap7} and supercontinuum generation in tapered step-index fibers \cite{tap2,tap3}. These studies will led us to consider the ultrashort beam propagation in tapered waveguide, as detailed in next section, modeled by dimensionless generalized HNLSE with complex potential
\begin{align}
    \no i~U_z+a_1(z)U_{xx}+a_2\vert U \vert^2 U+ i~[a_3(z)U_{xxx}&+a_4(\vert U \vert^2 U)_x +a_5U (\vert U \vert^2)_x]\\
    \label{vc} & +[V(z)-i~G(z)]U =0,
\end{align}
where $x,z,U$ are the dimensionless variables, $a_1(z)$ and $a_3(z)$ represents variable group velocity dispersion and third order dispersion,  respectively. The coefficient $a_2$ represents self-phase modulation, $a_4$ describes self-steepening effect and $a_5$ represents self-frequency shift arising from stimulated Raman scattering. Last complex term specify the inhomogeneous optical waveguide form with $V(z)$ and $G(z)$ as tapering and gain / loss functions which should be even and odd functions, respectively, for optical waveguide to be $\mathcal{PT}$-symmetric. For $V(z)=0$, Eq. (\ref{vc}) with variable coefficients is extensively studied to obtain soliton and rogue wave solutions \cite{hao,dai2012}.
Eq. (\ref{vc}) with $a_3(z)=0,a_4=0,a_5=0$, reduces to NLSE with complex potential, has been solved for multi-soliton solutions using Darboux transformation \cite{hao2}. In Ref. \cite{hp3}, authors have considered the Eq. (\ref{vc}) with variable coefficients and shown the existence of soliton-like solutions using Hirota bilinear method and symbolic computation. In this paper, we report exact self-similar solutions for Eq. (\ref{vc}), containing bright / dark similaritons, first- and second-order rogue waves, by reducing the generalized HNLSE into constant coefficient HNLSE using similarity transformation. The analytical results help to identify the allowed tapering profile in conjunction with appropriate gain function which manifested the complex potential to be $\mathcal{PT}$-symmetric. We demonstrate the controlling of ultrashort self-similar waves through judicious choice of free parameters.

\section{Generalization of variable coefficient HNLSE for tapered waveguide}
The ultrashort pulse propagation in inhomogeneous fiber / waveguide is governed by the dimensionless HNLSE with variable coefficients as follows \cite{hao,dai2012}
\be\label{eq1}
i q_\zeta+K_1'(\zeta ) q_{xx}+i K_2'(\zeta) q_{xxx}+a_2(\zeta )\vert q \vert^2 q+i a_4(\zeta ) (\vert q \vert^2q)_x+i a_5(\zeta ) q(\vert q \vert^2)_x=i \Gamma(\zeta)q,
\ee
where all the terms have their usual meaning as discussed earlier. For $\Gamma(\zeta)=0$, this model was first proposed by Papaioannou et al. for evolution of a femtosecond
duration pulse in a nonlinear optical fiber exhibiting a
small axial inhomogeneity \cite{Papaioannou}. For the case, when the axial inhomogeneity is small, the variations of
the dispersion terms of linear origin ($K_1', K_2'$) are
more significant than the nonlinear ones ($a_2, a_4, a_5$) \cite{Papaioannou}. Thus, the variation in the nonlinear terms can be ignored and treated as constants. In real fibers, losses produce attenuation and
broadening of the soliton. One method
to circumvent this problem is to taper the fiber core along axial direction \cite{Kazuhito}, so that the group-velocity dispersion (GVD) decreases with distance along the fiber
directly proportional to the exponential soliton attenuation
and inversely proportional to the square of the effective core
radius. Therefore, the variable dispersion terms can be modified as
\be\no
K_1'(\zeta)=\frac{K_1(\zeta)}{R^2(\zeta)};~~~~K_2'(\zeta)=\frac{K_2(\zeta)}{R^2(\zeta)},
\ee
where $R(\zeta)$ is the effective core radius, $K_1(\zeta)$ and $K_2(\zeta)$ are the variable dispersion coefficients.
Using transformation
\be\no
q(x,\zeta)=A(\zeta) U(x,z) e^{i k \zeta},
\ee
Eq. (\ref{eq1}) reduces to
\begin{align}
    \no i U_z+\frac{K_1(\zeta)}{A^2(\zeta)R^2(\zeta)}U_{xx}+i \frac{K_2(\zeta)}{A^2(\zeta)R^2(\zeta)}U_{xxx} +a_2 \vert U \vert^2 U&+
    i a_4(\vert U \vert^2 U)_x+ia_5 U(\vert U \vert^2)_x\\
    \label{eq2}&=\frac{k}{A^2}U+i \(\frac{\Gamma(\zeta)A-A'}{A^3}\)U,
\end{align}
such that $z(\zeta)=\int_{0}^{\zeta}A^2(\zeta)d\zeta$. Assuming $A^2(\zeta)=\frac{1}{R^2(\zeta)}$,
Eq. (\ref{eq2}) reads
\begin{align}
    \no iU_z+K_1(\zeta)U_{xx}+i K_2(\zeta)U_{xxx} & +  a_2 \vert U \vert^2 U+i a_4(\vert U \vert^2 U)_x+i a_5 U(\vert U \vert^2)_x\\
    \label{eq3} &=kR^2(\zeta)U+i\(\Gamma(\zeta)R^2(\zeta)+\frac{R'(\zeta)}{R(\zeta)}\)U.
\end{align}
 Under the coordinate transformation $\zeta\rightarrow z$, the variable terms can be identified as
$K_1(\zeta)=a_1(z),K_2(\zeta)=a_3(z),R^2(\zeta)=-\frac{V(z)}{k},\Gamma(\zeta)R^2(\zeta)+\frac{R'(\zeta)}{R(\zeta)}=G(z)$, and the Eq. (\ref{eq3})
reduces to the model equation, given by Eq. (\ref{vc}), under consideration in this paper. Here, $V(z)$ comes out to be directly proportional to the square of the effective core radius, known as taper cross-section or specifically tapering profile.

\section{Similarity transformation and self-similar solutions} 
\begin{figure}[h!]
    \begin{center}
        \subfigure[]{
        \includegraphics[scale=0.56]{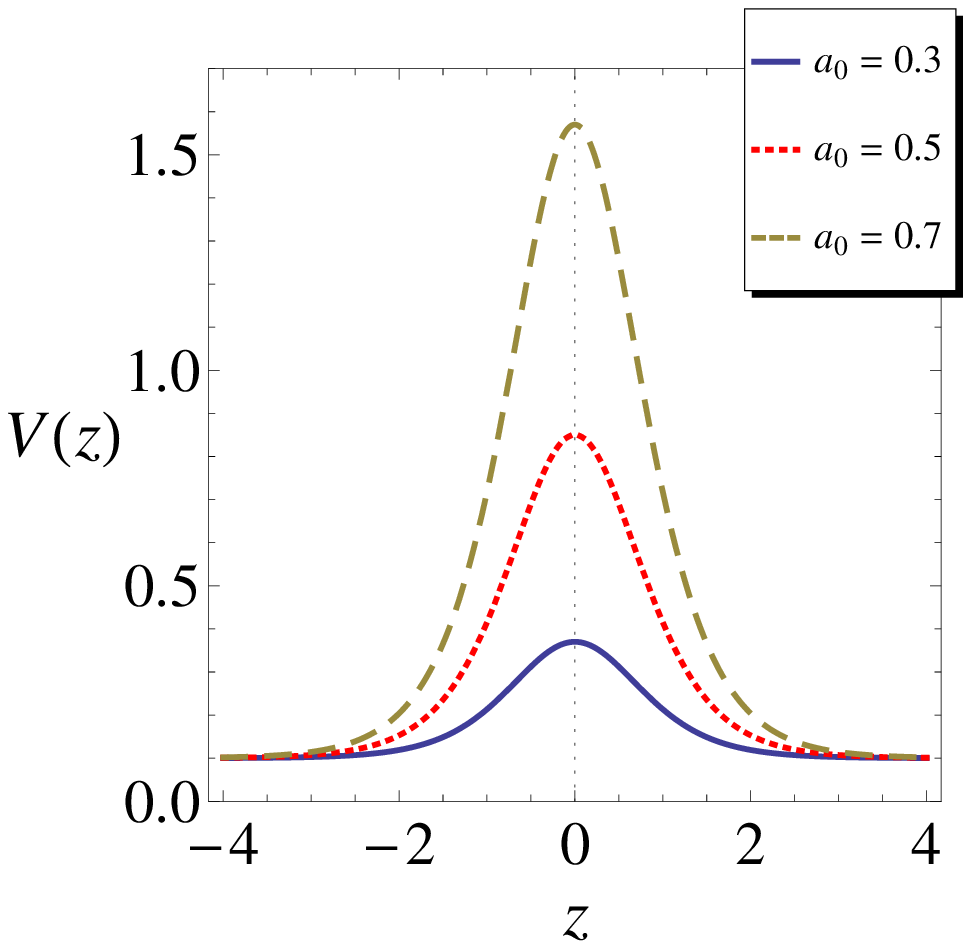}
        }
        \subfigure[]{
        \includegraphics[scale=0.55]{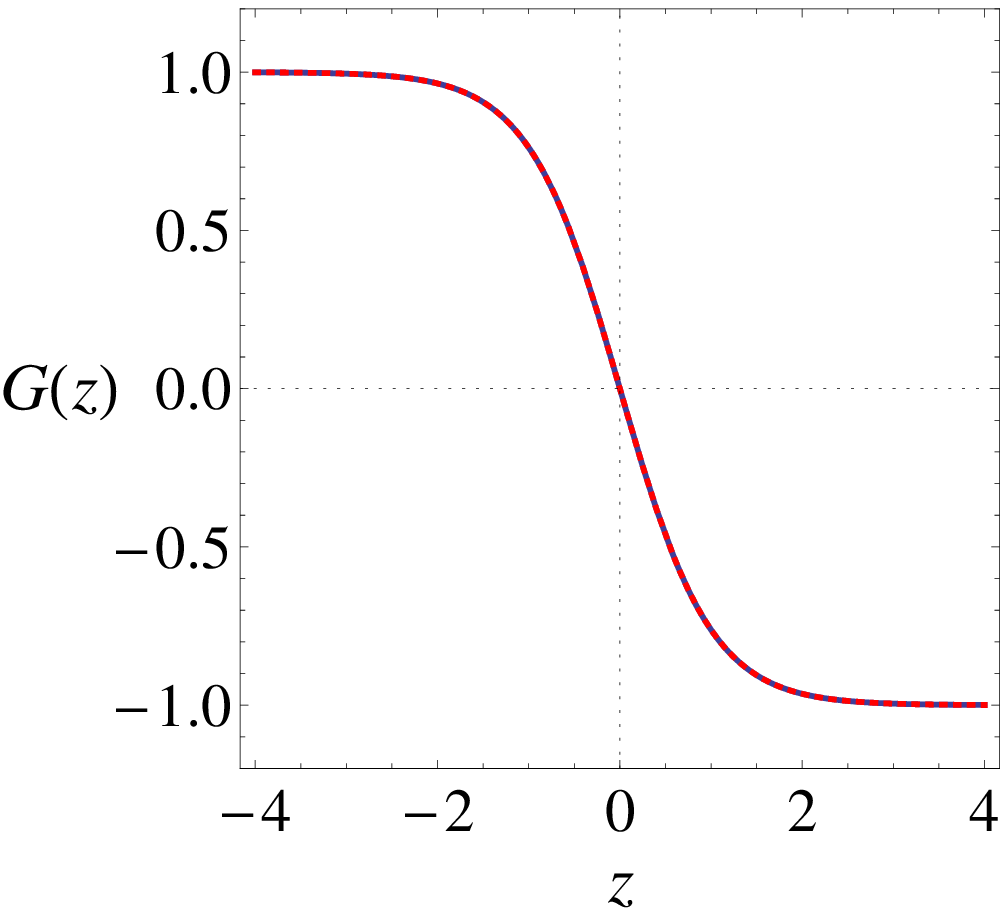}
        }
        \caption{\label{taper} (a) Tapering profile for different values of $a_0$, and (b) gain / loss profile. The
        values of other parameters used in the plots are mentioned in the text.}
    \end{center}
    \end{figure}

The exact solution of Eq. (\ref{vc}) can be obtained by using following gauge and similarity transformation \cite{pono1}
    \be \label{ansatz}U(x,z)=A(z)~\psi[\chi(x,z),\xi(z)]~e^{i \phi(x,z)}, \ee
where $A(z)$ is the amplitude of self-similar wave, $\chi(x,z)$ and $\xi(z)$ are self-similar variables such that $\chi(x,z)=x-x_c(z)$ with $x_c(z)$ corresponds to the position of the self-similar wave center. The phase is given as $\phi(x,z)=\gamma x+\delta z$ where $\gamma$ and $\delta$ are parameters related to frequency shift and the phase offset, respectively. Substituting Eq. (\ref{ansatz}) into Eq. (\ref{vc}), we obtain constant coefficient HNLSE
    \be \label{hnls} i\psi_\xi+b_1\psi_{\chi\chi}+(a_2-a_4\gamma)\vert\psi\vert^2\psi+i[b_3\psi_{\chi\chi\chi}
    +a_4(\vert\psi\vert^2\psi)_\chi+a_5\psi(\vert\psi\vert^2)_\chi]=0, \ee
such that group velocity dispersion, third order dispersion, effective propagation distance and guiding center position, respectively, are given by
    \be\no a_1(z)=\left(b_1+3b_3 \gamma\right)A^2(z),~~~a_3(z)=b_3A^2(z),~~~\xi(z)=\xi_0+\int_{0}^{z}A^2(z)dz,\ee
    \be\no x_c(z)=x_0+\gamma\left(2b_1+3b_3 \gamma\right)\int_{0}^{z}A^2(z)dz,\ee
with $x_c(0)=x_0$, $\xi(0)=\xi_0$. Further, tapering and gain / loss functions reads
    \be\no  V(z)=\delta+ \gamma^2\left(b_1+2b_3 \gamma \right)A^2(z),~~~ G(z)=\frac{1}{A(z)} \frac{dA(z)}{dz}. \ee
Here, amplitude function $A(z)$ can be chosen arbitrarily, pertaining to the condition that inhomogeneous optical waveguide should be $\mathcal{PT}$-symmetric. For $A(z)=a_0~\mbox{sech}(z)$, where $a_0$ is free parameter, tapering and gain / loss function is given as
  \be\label{real} V(z)= \delta+ {a_0}^2~\gamma^2\left(b_1+2b_3 \gamma \right)\mbox{sech}^2(z),~~~ G(z)=-\tanh(z). \ee
This type of tapering profile follows from the theory of sech$^2$-profile waveguide in optics \cite{lamb}. In recent years, a significant work has been done to study the evolution of soliton-like solutions in sech$^2$-type tapered waveguides \cite{pono1,pra1}. Here, the tapering profile can be modulated through various parameters while the gain profile is independent of modulations in the tapering profile. The parameter `$\delta$' signifies the uniform profile of waveguide along longitudinal direction. We have shown the tapering profiles for different values of $a_0$ in Fig. \ref{taper}(a) and corresponding gain / loss profile in Fig. \ref{taper}(b). Here, the tapering amplitude increases as value of $a_0$ increases, so we can termed $a_0$ as tapering parameter.  The other parameters used are $b_1=1,b_3=1,\delta=0.1$ and $\gamma=1$. From plots, one can observe that tapering function $V(z)$ is an even-function whereas gain / loss function $G(z)$ is an odd-function of position which corresponds to $\mathcal{PT}$-symmetric waveguide.

\par It is well known that constant coefficient HNLSE, given by Eq. (\ref{hnls}), is exactly solved and has localized solutions, given by bright / dark solitons \cite{bd2,bd3} and rogue waves \cite{rw3} for specific choice of model parameters. As stated earlier, similarity transformation Eq. (\ref{ansatz}), establishes one to one correspondence between generalized HNLSE and constant coefficient HNLSE. For all the localized solutions of constant coefficient HNLSE, the corresponding optical self-similar solutions of Eq. (\ref{vc}) can be obtained by means of the reverse transformation variables and functions. In next section, we will study the evolution of self-similar solutions for tapering profile given by Eq. (\ref{real}).

 \begin{figure}[h!]
    \begin{center}
    \subfigure[]{
    \includegraphics[scale=0.58]{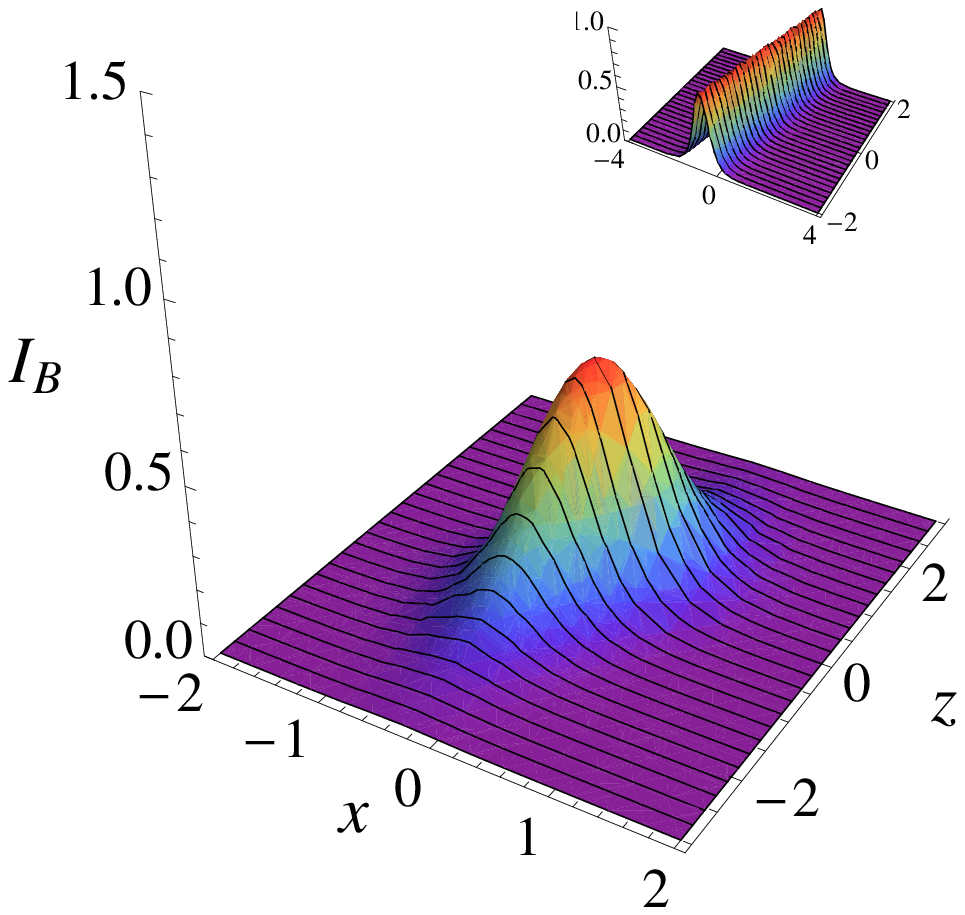}
    }
    \subfigure[]{
    \includegraphics[scale=0.53]{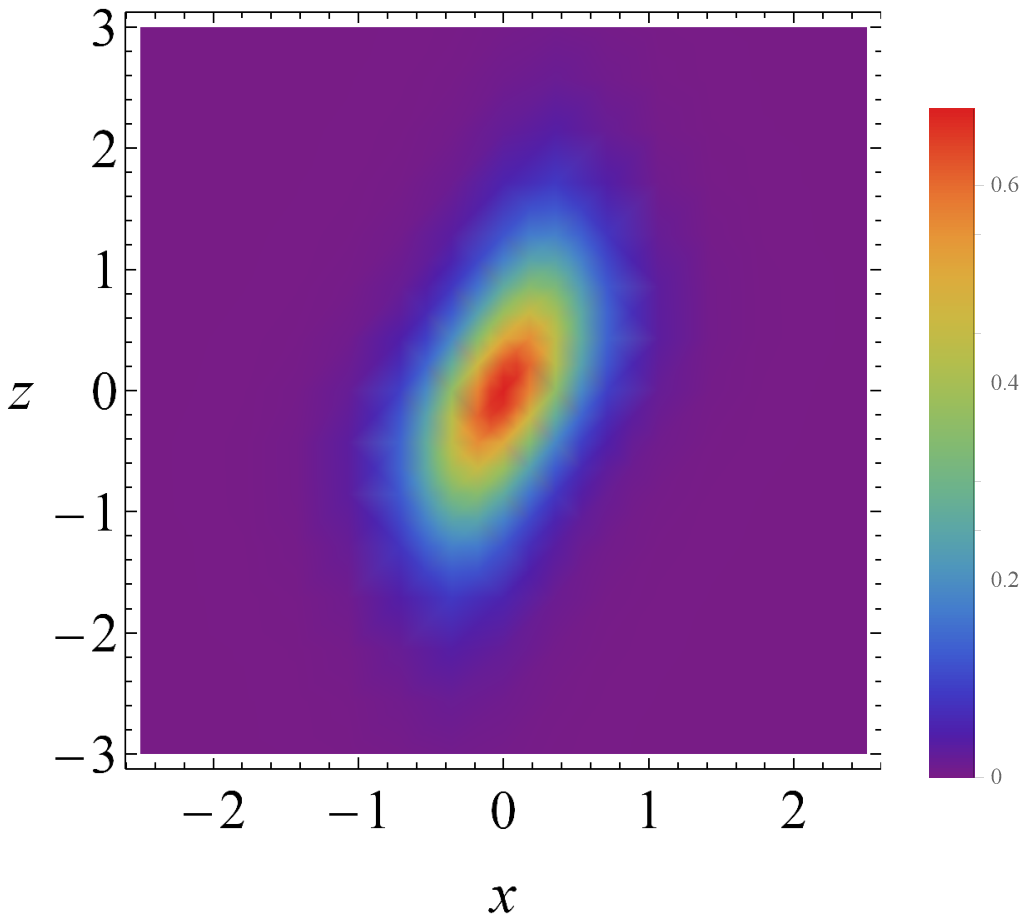}
    }
    \caption{\label{bright-int} (a) Intensity profile of bright similariton, and (b) corresponding contour plot for $A(z)=a_0~\mbox{sech}(z)$ with $a_0=0.5$. (Inset) The intensity profile of bright similariton for non-tapered waveguide with $A(z)=a_0$. The values of other parameters used in the plots are mentioned in the text.}
    \end{center}
    \end{figure}

    \begin{figure}
    \begin{center}
    \subfigure[]{
    \includegraphics[scale=0.55]{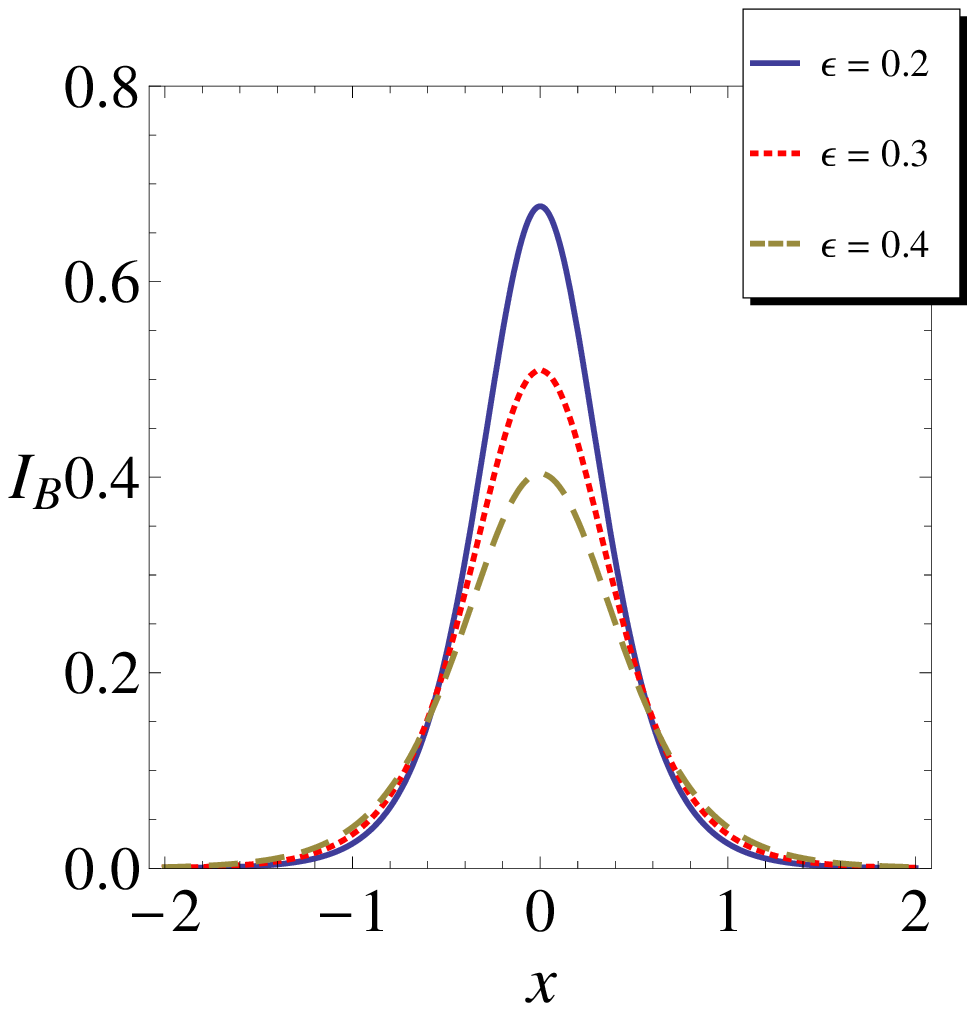}
    }
    \subfigure[]{
    \includegraphics[scale=0.55]{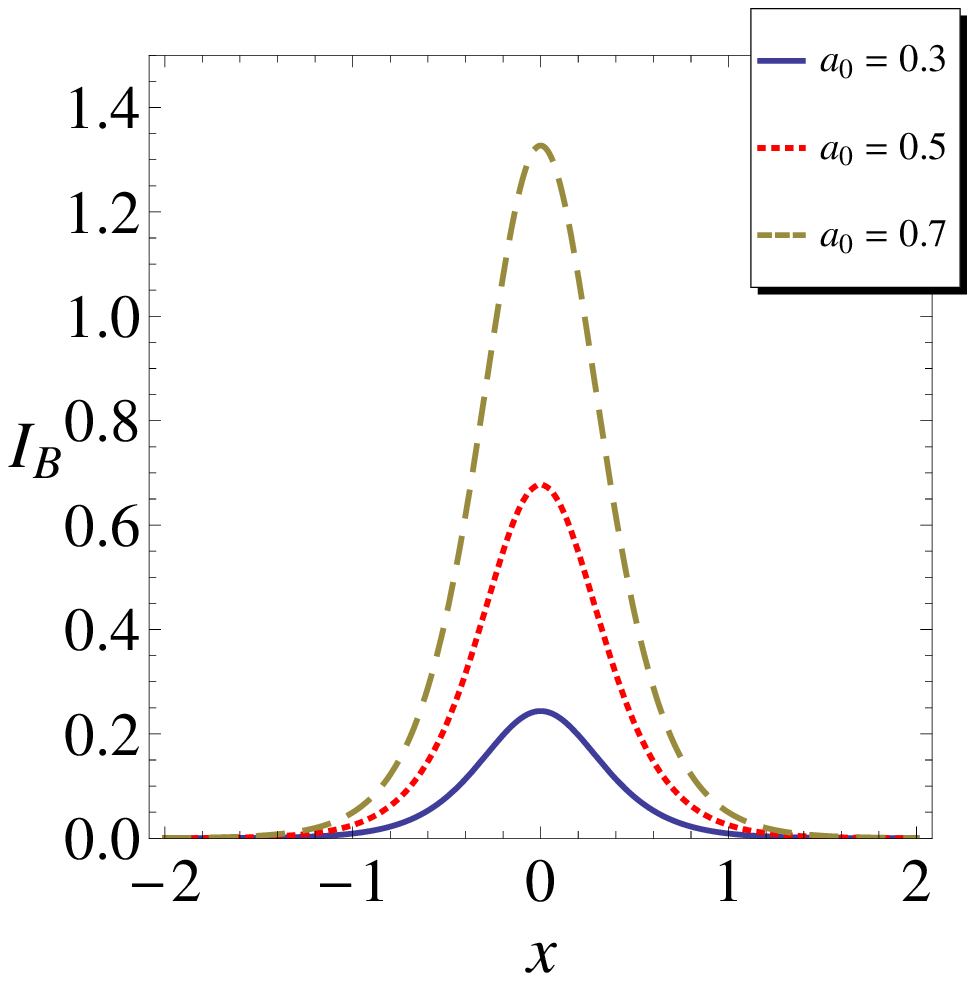}
    }
    \subfigure[]{
    \includegraphics[scale=0.55]{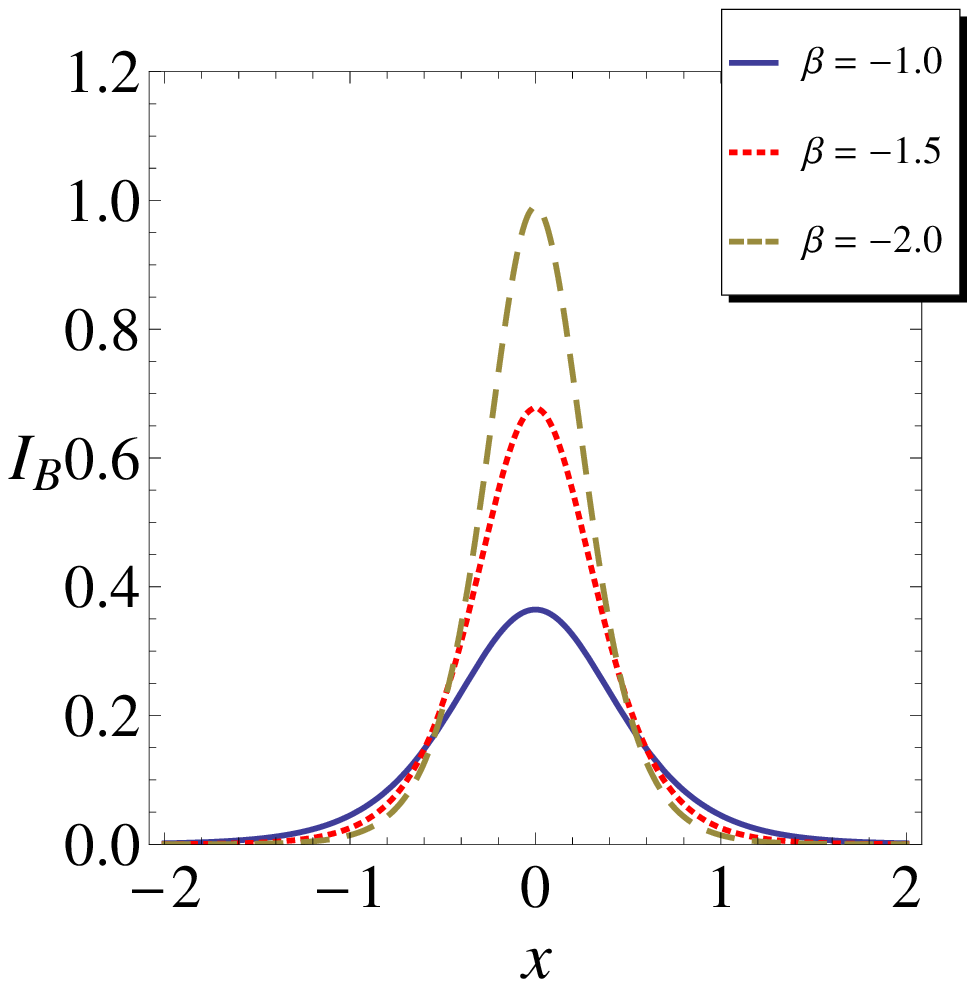}
    }
    \caption{ \label{bright} (a,b,c) Effect of $\epsilon,a_0$ and $\beta$ on the intensity of bright similariton at $z=0$.}
    \end{center}
    \end{figure}

\subsection{Bright and dark similaritons}
Substituting $\psi(\chi,\xi)=\rho(\zeta) e^{i(k \xi-\omega  \chi)}$, where $\zeta=\chi+\beta~\xi$ and $\beta,k,\omega$ are real parameters, in Eq. (\ref{hnls}) and solving the coupled equations for $\omega=\frac{3 a_2 b_3-2 a_5 b_1-3 a_4 \left(b_1+\gamma b_3\right)}{6 \left(a_4+a_5\right) b_3}$ and $k=\frac{2 \omega  b_1^2-3 \beta  \omega  b_3+8 \omega ^2 b_1 b_3+8 \omega ^3 b_3^2-\beta  b_1}{b_3}$, we obtain
    \be \label{int} \rho_{\zeta \zeta}+p\rho+ q \rho^3=0,\ee
with $p=\left(\frac{\beta -3b_3\omega ^2 -2 b_1 \omega }{b_3}\right)$ and $q=\left(\frac{3 a_4+2 a_5}{3 b_3}\right)$. For $p<0$ and $q>0$, Eq.(\ref{int}) possesses bright solitons and the corresponding soliton solution for constant coefficient HNLSE can be written as \cite{bd2}
    \be\no \psi(\chi,\xi)= \sqrt{-\frac{2p}{q}}~\mbox{sech}\(\sqrt{-p}\(\chi+\beta~\xi\)\)~e^{i(k \xi-\omega  \chi)}. \ee
Under the conditions $p>0$ and $q<0$, for dark solitons of Eq. (\ref{int}), the soliton solution for constant coefficient HNLSE reads \cite{bd3}
    \be\no \psi(\chi,\xi)= \sqrt{-\frac{p}{q}}~\mbox{tanh}\left(\sqrt{\frac{p}{2}}\(\chi+\beta~\xi\)\right)~e^{i(k \xi-\omega  \chi)}. \ee
The corresponding self-similar solutions of Eq. (\ref{vc}) can be
obtained by means of the reverse transformation variables and
functions. The general expression of intensity, $|U(x,z)|^2=A^2|\psi|^2$, for bright similariton $I_B$ and dark similariton $I_D$ is given as
    \begin{align}
     \label{int1}I_B(x,z)&= \(-\frac{2p}{q}\)A^2(z)~\text{sech}^2\(\sqrt{-p}\(\chi+\beta~\xi\)\),\\
     \label{int2}I_D(x,z)&= \(-\frac{p}{q}\)A^2(z)~\text{tanh}^2\left(\sqrt{\frac{p}{2}}\(\chi+\beta~\xi\)\right).
    \end{align}
To study the evolution of similaritons, we consider the model parameters corresponds to integrable Sasa-Satsuma equation \cite{ssa}, such as $b_1=\frac{1}{2},\gamma=\frac{a_2-1}{a_4},b_3=\epsilon,a_4= 6\epsilon$ and $a_5=-3\epsilon$ in Eq. (\ref{hnls}). For these choices, the parameter $q$ can be obtained as $q=\left(\frac{3 a_4+2 a_5}{3 b_3}\right)=4$ which leads to the possibility of bright solitons for Eq. (\ref{hnls}), pertaining to the condition $\beta<-\frac{0.0833}{\epsilon}$ for $a_2=0.8$. It means for positive values of $\epsilon$, the wave parameter `$\beta$' can take only negative values where sign of $\beta$ attributes to the direction of propagation of bright similariton. In Fig. \ref{bright-int}, we have depicted the intensity profile (given by Eq. (\ref{int1})) and the corresponding contour plot of bright similariton for $\epsilon =0.2,\beta =-1.5$ and $A(z)=a_0~\mbox{sech}(z)$ with $a_0=0.5$. One can observe from the plots that intensity profile reveals the self-similar behavior, that is maintaining the shape but amplitude and width is changing, compared to the constant intensity evolution for non-tapered waveguide (shown in the inset of Fig. \ref{bright-int}(a) for $A(z)=a_0$). The intensity profile is confined to the tapering region  due to the choice of amplitude function $A(z)$ and asymptotically goes to zero along both sides of the tapering region. The contour plot clearly depicts the confinement of bright similaritons along the tapered region in $xz-$plane and variation of intensity from zero to maximum value at centre of the tapering region. The intensity of bright similariton can be controlled through various model and wave parameters. First, we analyse the effect of higher-order terms on intensity of bright similariton and present the sectional plots of the intensity variation with parameters `$\epsilon$' in Fig. \ref{bright}(a). One can observe that maximum intensity of similariton decreases as impact of higher-order terms increases. Figs. \ref{bright}(b,c) represents the effect of tapering parameter `$a_0$' and wave parameter `$\beta$', respectively, and in both cases, it is found that intensity of similariton can be increased with increase in tapering amplitude and magnitude of wave parameter `$\beta$'.

 \begin{figure}[h!]
    \begin{center}
    \subfigure[]{
    \includegraphics[scale=0.53]{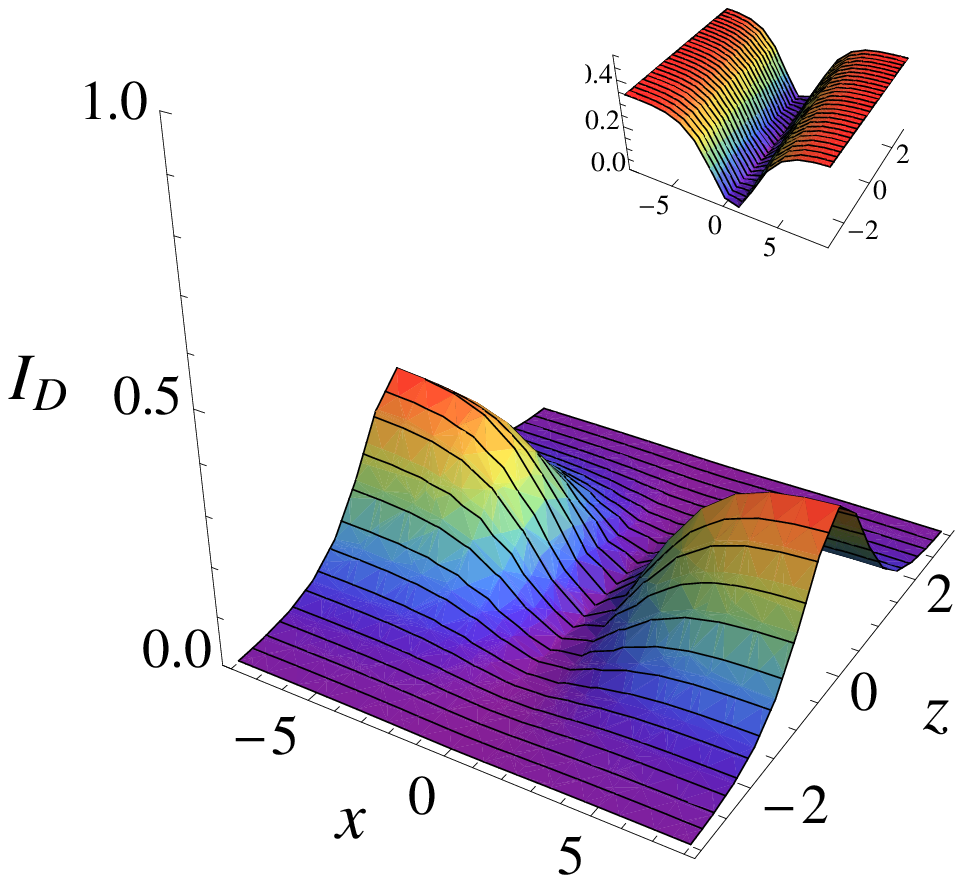}
    }
    \subfigure[]{
    \includegraphics[scale=0.48]{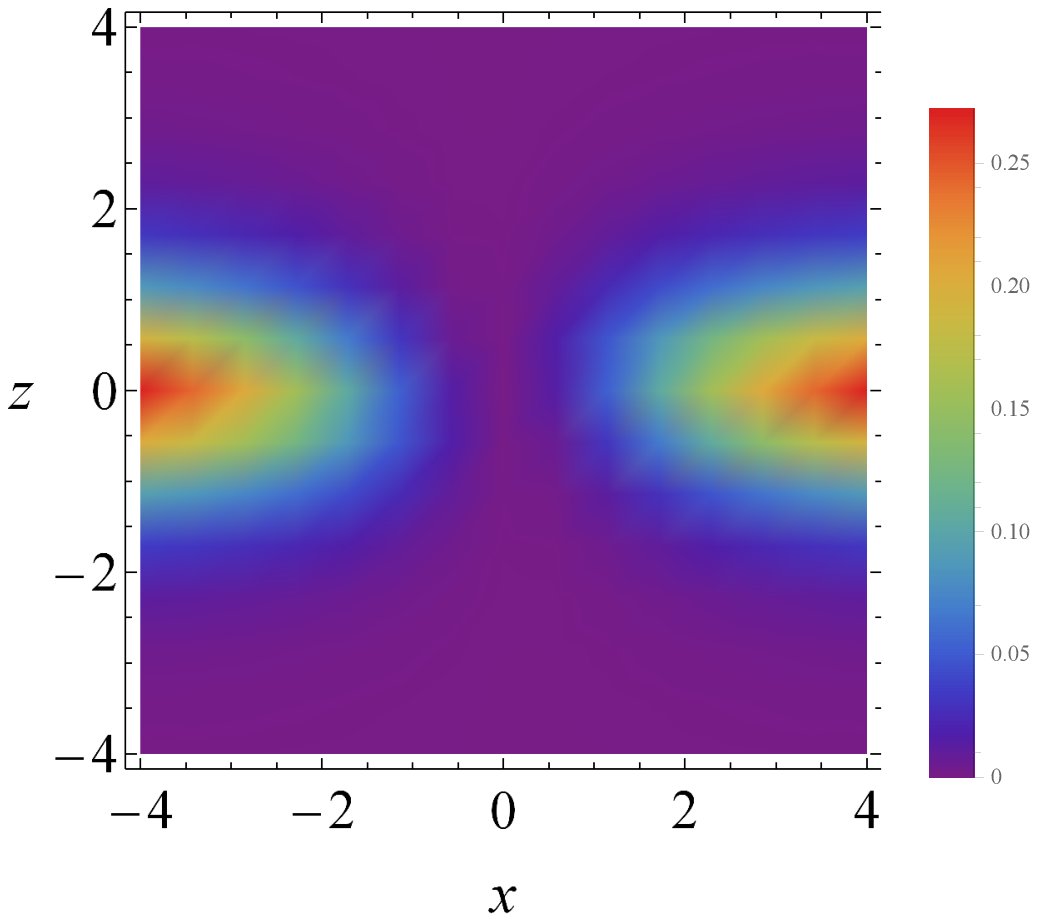}
    }
    \caption{\label{dark-int} (a) Intensity profile of dark similariton, and (b) corresponding contour plot for $A(z)=a_0~\mbox{sech}(z)$ with $a_0=0.5$. (Inset) The intensity profile of dark similariton for non-tapered waveguide with $A(z)=a_0$. The values of other parameters used in the plots are mentioned in the text.}
    \end{center}
\end{figure}

\begin{figure}[h!]
    \begin{center}
    \subfigure[]{
    \includegraphics[scale=0.55]{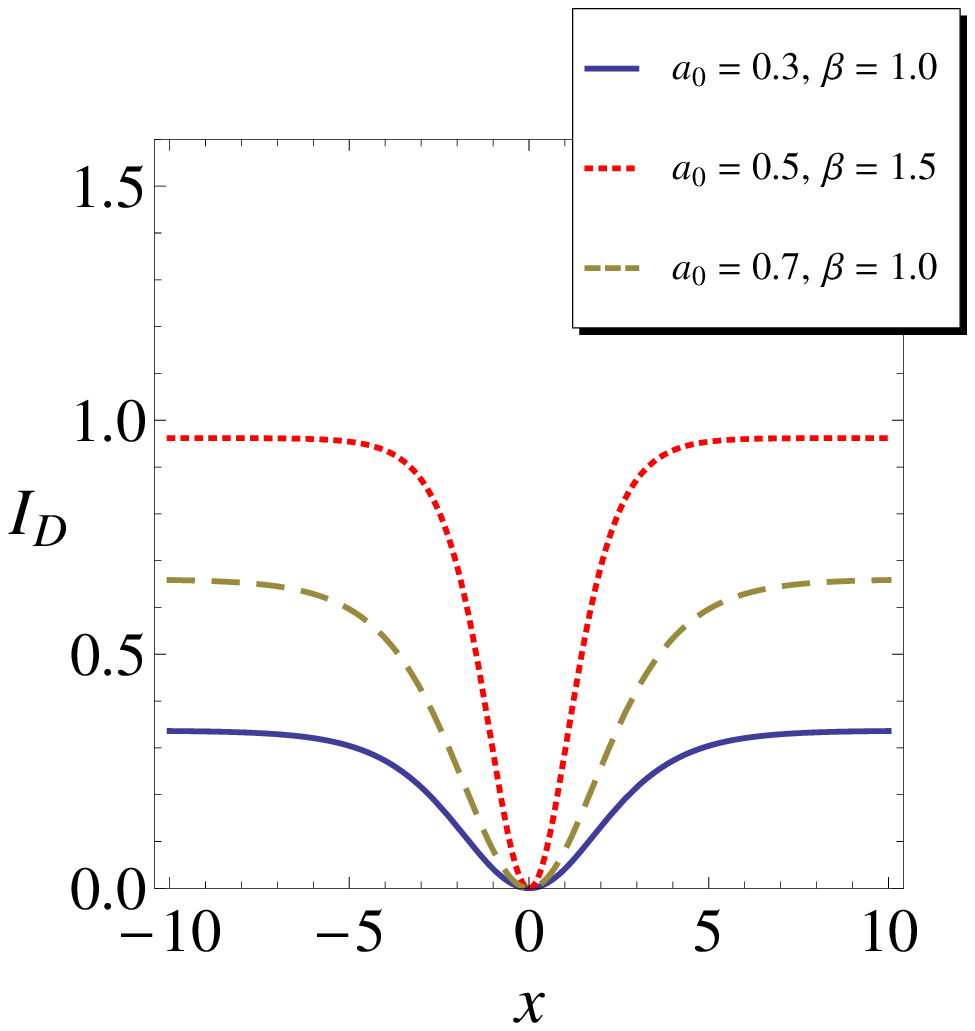}
    }
    \subfigure[]{
    \includegraphics[scale=0.55]{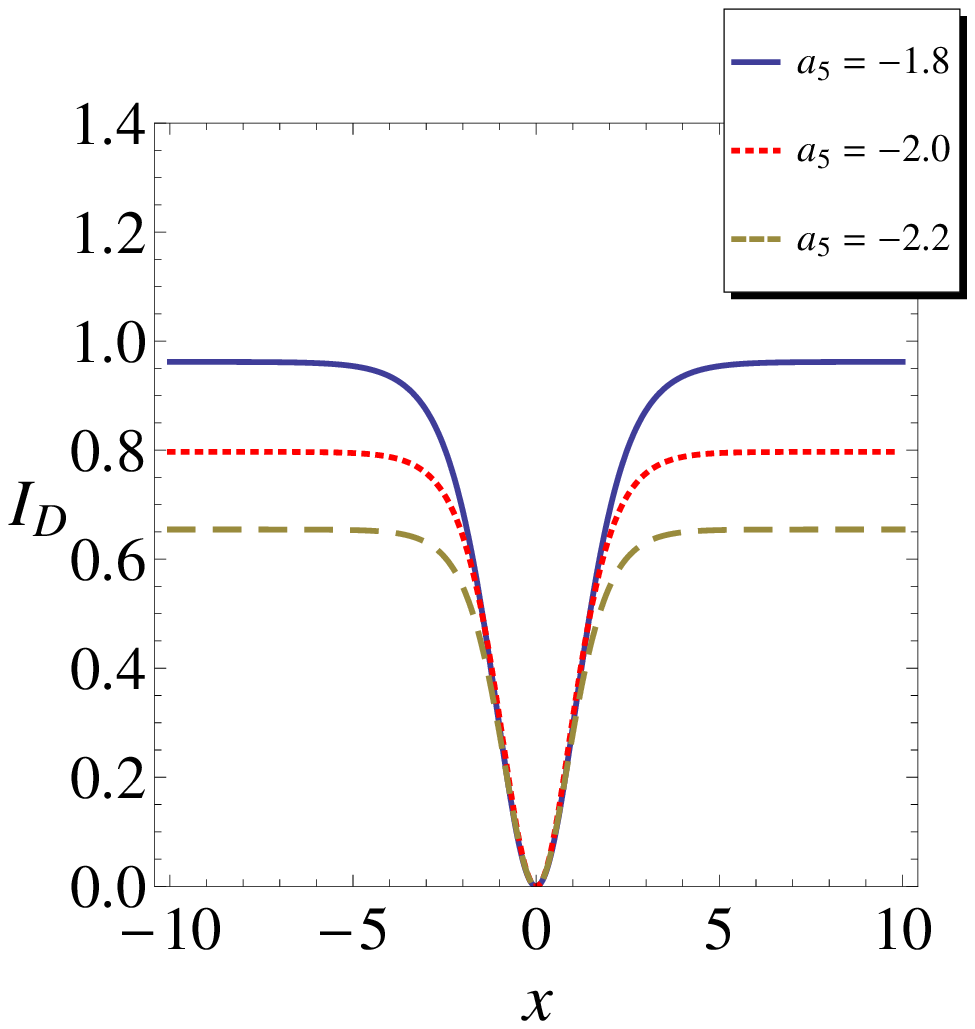}
    }
    \caption{ \label{dark} Effect of (a) $a_0,\beta$, and (b) self-frequency shift parameter `$a_4$' on the intensity of dark similariton at $z=0$.}
    \end{center}
 \end{figure}

\par The evolution of dark similariton is studied for typical values of the model parameters as $b_1=\frac{1}{2},a_2=0.8,b_3=1,a_4=1,a_5=-1.8$ and assuming $\gamma=\frac{a_2-1}{a_4}$. For these values, the dark soliton exists for Eq. (\ref{hnls}) with the condition $\beta>0.730$. In Fig. \ref{dark-int}, we have shown the intensity profile (given by Eq. (\ref{int2})) and the corresponding contour plot of dark similariton for $\beta =1$ and $A(z)=a_0~\mbox{sech}(z)$ with $a_0=0.5$. Like bright similariton, the intensity of dark similariton also asymptotically goes to zero along both sides of the tapering region compared to the constant intensity profile for non-tapered waveguide (shown in the inset of Fig. \ref{dark-int}(a) for $A(z)=a_0$). In Fig. \ref{dark}(a), it is shown that maximum intensity of dark similariton increases as amplitude of $a_0$ or $\beta$ increases. Fig. \ref{dark}(b) represents sectional plot for the effect of self-frequency shift parameter on the intensity of similariton and one can observe that intensity decreases as magnitude of `$a_5$' increases. Further, it is observed that intensity of dark similariton increases as the positive value of self-steepening parameter `$a_4$' increases.

 \begin{figure}[h!]
    \begin{center}
    \subfigure[]{
    \includegraphics[scale=0.53]{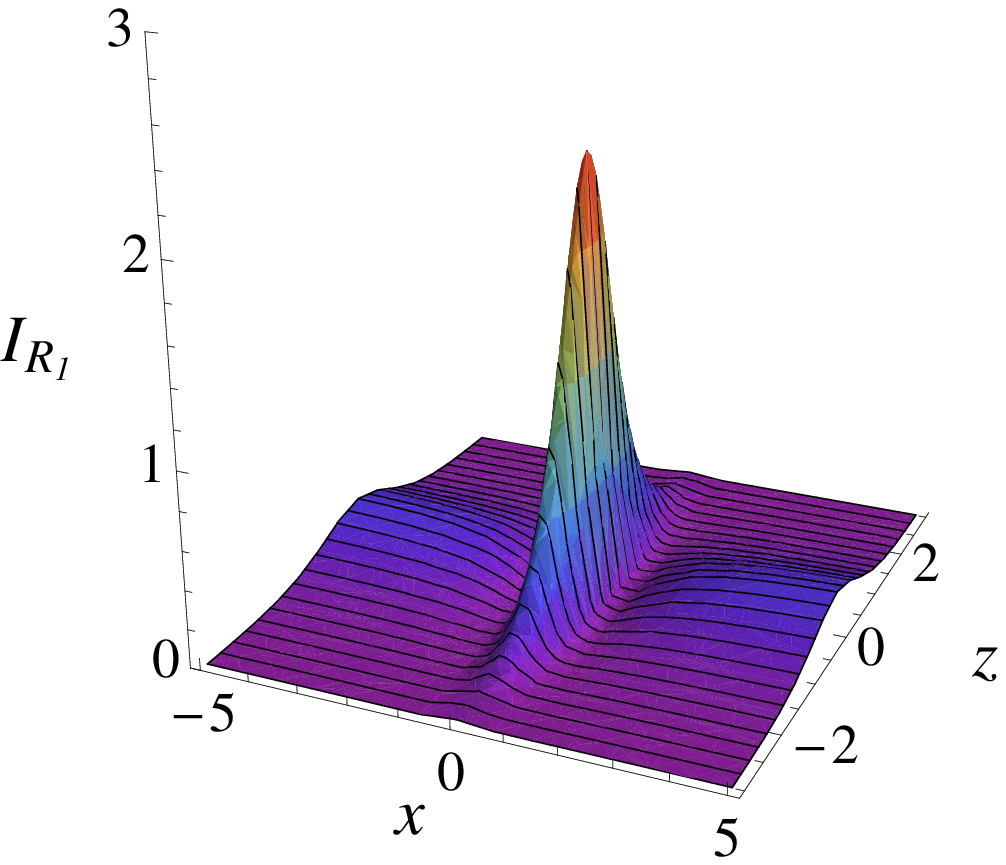}
    }
    \subfigure[]{
    \includegraphics[scale=0.53]{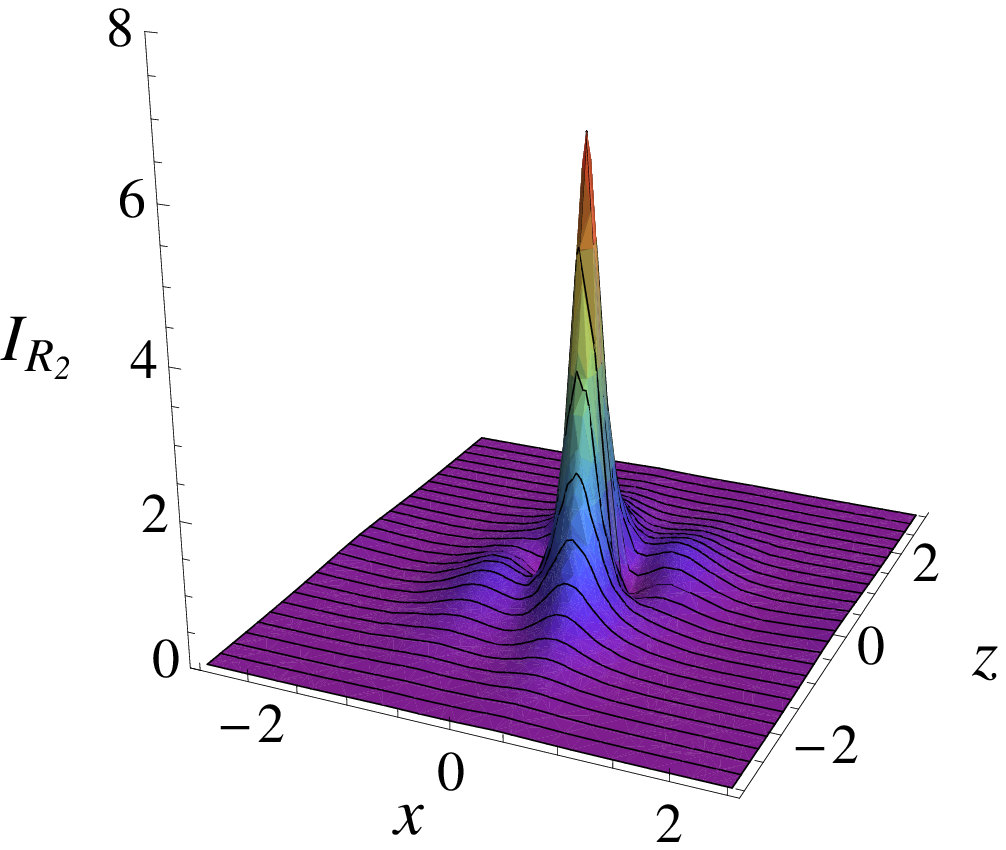}
    }
    \caption{\label{rogue1-int1} Intensity profile of (a) self-similar first-order and (b) second-order rogue waves for $A(z)=a_0~\mbox{sech}(z)$ with $a_0=0.5$. The values of other parameters used in the plots are mentioned in the text.}
    \end{center}
 \end{figure}

\subsection{Self-similar rogue wave solutions}
Ankiewicz et al. \cite{rw3} presented the explicit form for first- and second-order rogue wave solutions (rational solutions) of Hirota equation employing Darboux transformation technique. For $b_1=\frac{1}{2}, \gamma=\frac{a_2-1}{a_4}, b_3=-\alpha, a_4=-6\alpha, a_5=6\alpha$, Eq. (\ref{hnls}) reduces to integrable Hirota equation, for which first-order rogue wave solution is given as
    \be\no \psi(\chi,\xi)=-\left(1-4\frac{1+2i \xi  }{1+4(\chi +6\alpha  \xi )^2+4\xi ^2}\right)e^{i \xi}, \ee
and second-order rogue wave solution reads
    \be\no \psi(\chi,\xi)=\left(1+\frac{G+i \xi  H}{F}\right)e^{i \xi}, \ee
where
     {\small \begin{align}
   \no G & = 12(-16\chi^4-384 \alpha  \xi \chi^3-24\left(4\left(36\alpha ^2+1\right)\xi^2+1\right)\chi^2-96\alpha
    \xi\left(12\left(12\alpha ^2+1\right)\xi^2+7\right)\chi \\ &~~~\no
    -16\left(1296\alpha ^4+216\alpha ^2+5\right)\xi^4-72\left(44\alpha ^2+1\right)\xi^2+3),\\
   \no H & = 24(-16\chi^4-384\alpha \xi \chi^3-8\left(\left(432\alpha ^2+4\right)\xi^2-3\right)\chi^2-96\alpha  \xi \chi\left(4\left(36\alpha ^2+1\right)\xi^2+1\right)  \\ &~~~\no
    -16\left(36\alpha ^2+1\right)^2\xi^4-8\left(180\alpha ^2+1\right)\xi^2+15),\\
   \no F & = 64\chi^6+2304 \alpha  \xi \chi^5-432\left(624 \alpha ^4-40 \alpha ^2-1\right)\xi^4+36\left(556 \alpha ^2+11\right)\xi^2+9
    +64\left(36 \alpha ^2+1\right)^3\xi^6 \\ &~~~\no
    +384 \alpha  \chi^3\left(12\left(60 \alpha^2+1\right)\xi^2-1\right)\xi+48\chi^4\left(\left(720 \alpha ^2+4\right)\xi^2+1\right)
    +12\left(16(6480 \alpha ^4+216 \alpha ^2+1\right)\xi^4  \\ &~~~\no
    -24\left(60 \alpha ^2+1\right)\xi^2+9)\chi^2+144 \alpha \xi(16\left(36 \alpha ^2+1\right)^2\xi^4
    +\left(8-864 \alpha ^2\right)\xi^2+17)\chi.
    \end{align}}

\begin{figure}[h!]
    \begin{center}
    \subfigure[]{\includegraphics[scale=0.52]{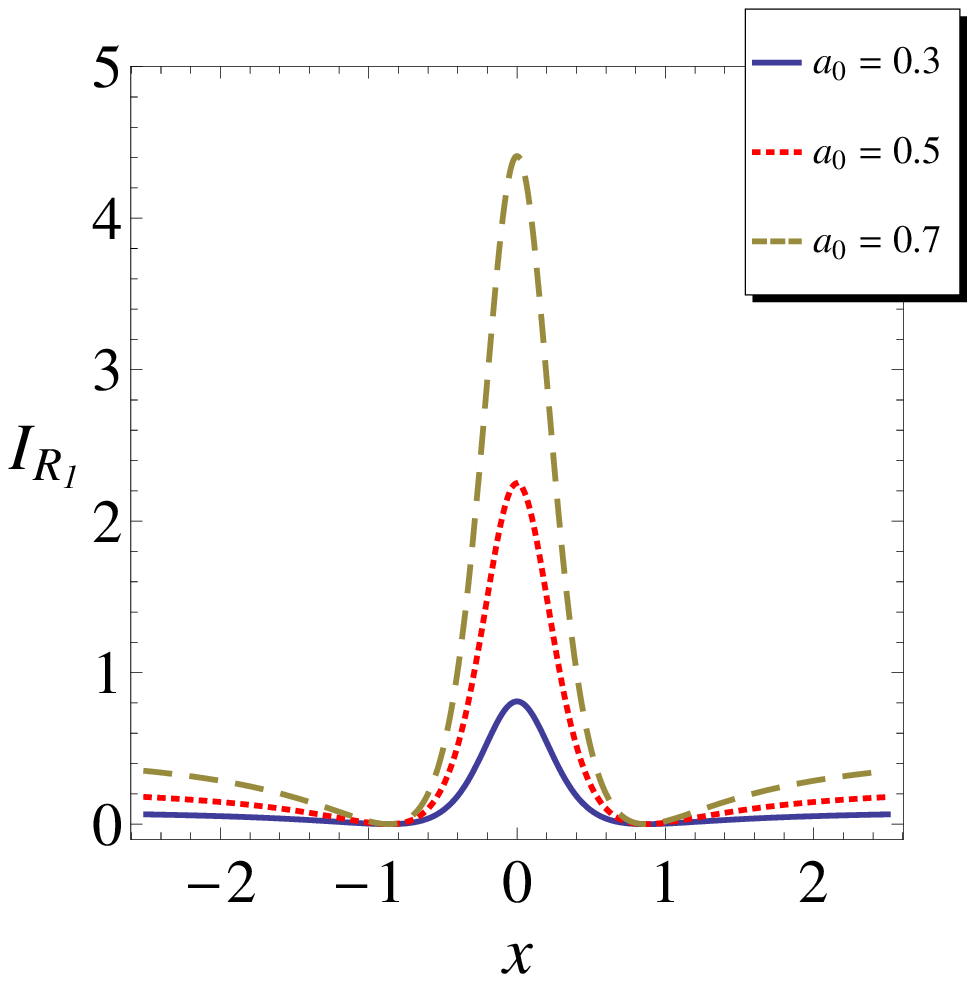}}
    \subfigure[]{\includegraphics[scale=0.54]{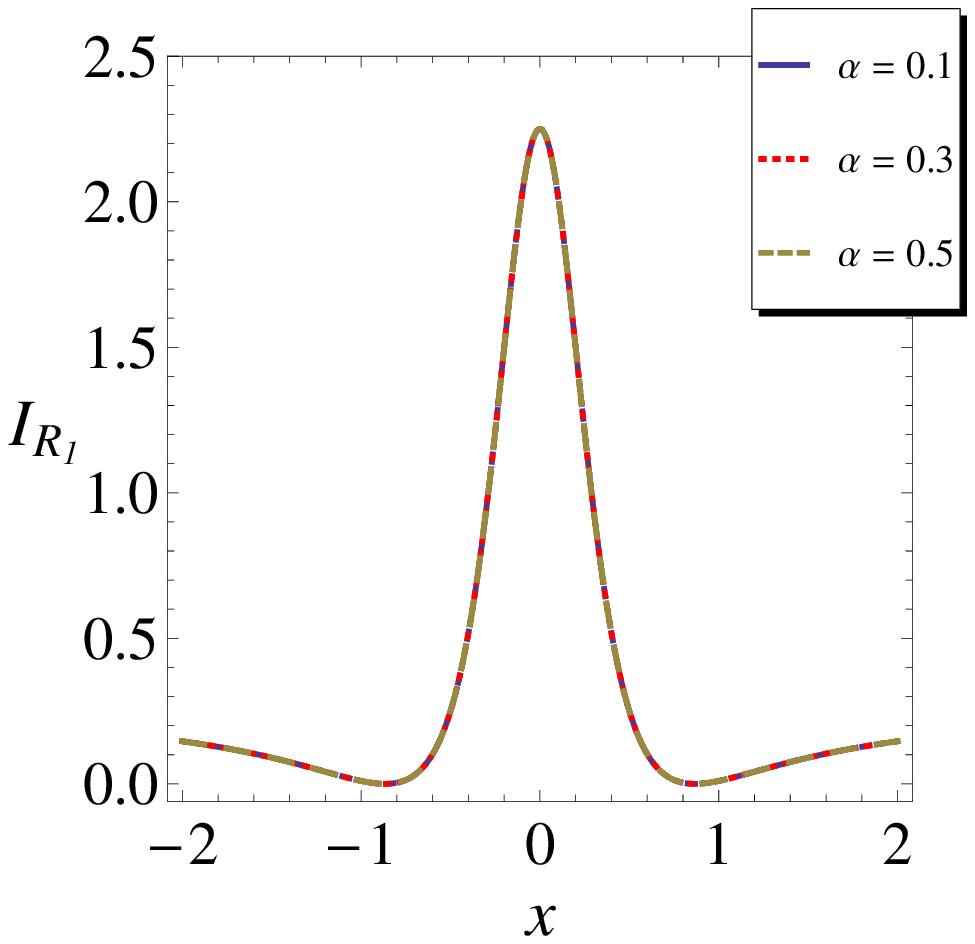}}
    \subfigure[]{\includegraphics[scale=0.54]{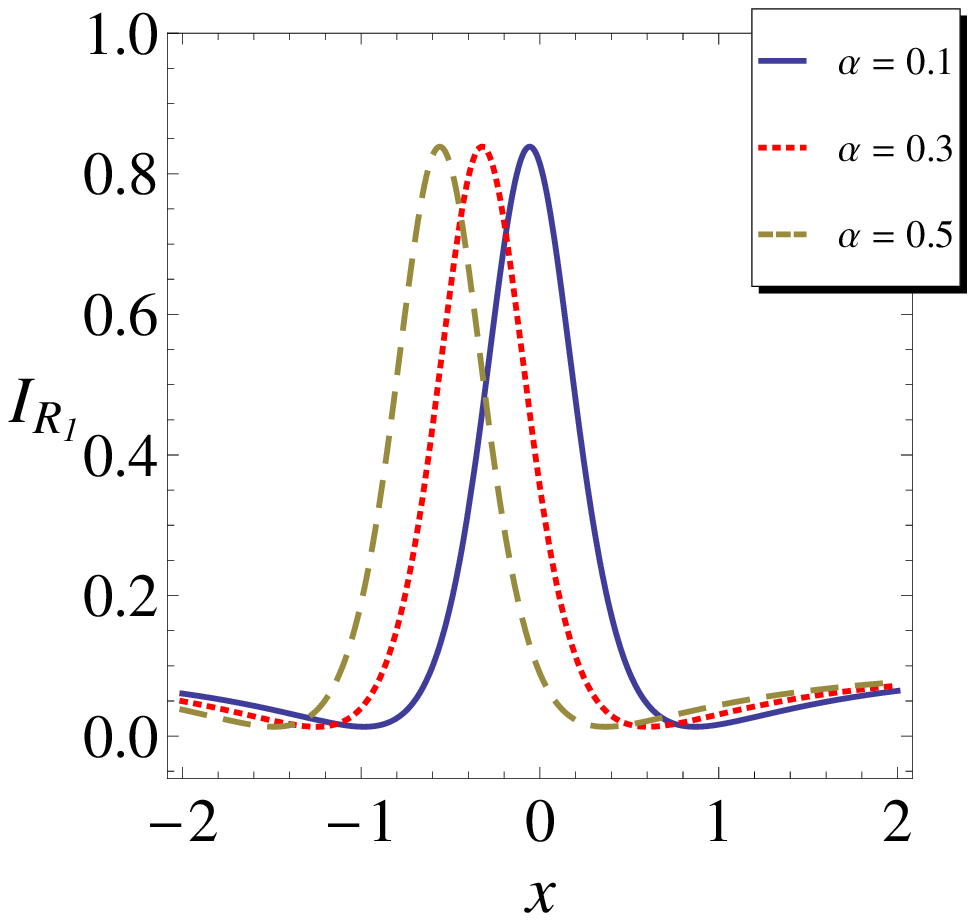}}
    \caption{\label{rogue1} Effect of (a) $a_0$ at $z=0$, and (b,c) higher-order parameter `$\alpha$' at $z=0$ and $z=1$, respectively, on the intensity of self-similar first-order rogue waves.}
    \end{center}
\end{figure}

\begin{figure}[h!]
    \begin{center}
    \subfigure[]{\includegraphics[scale=0.57]{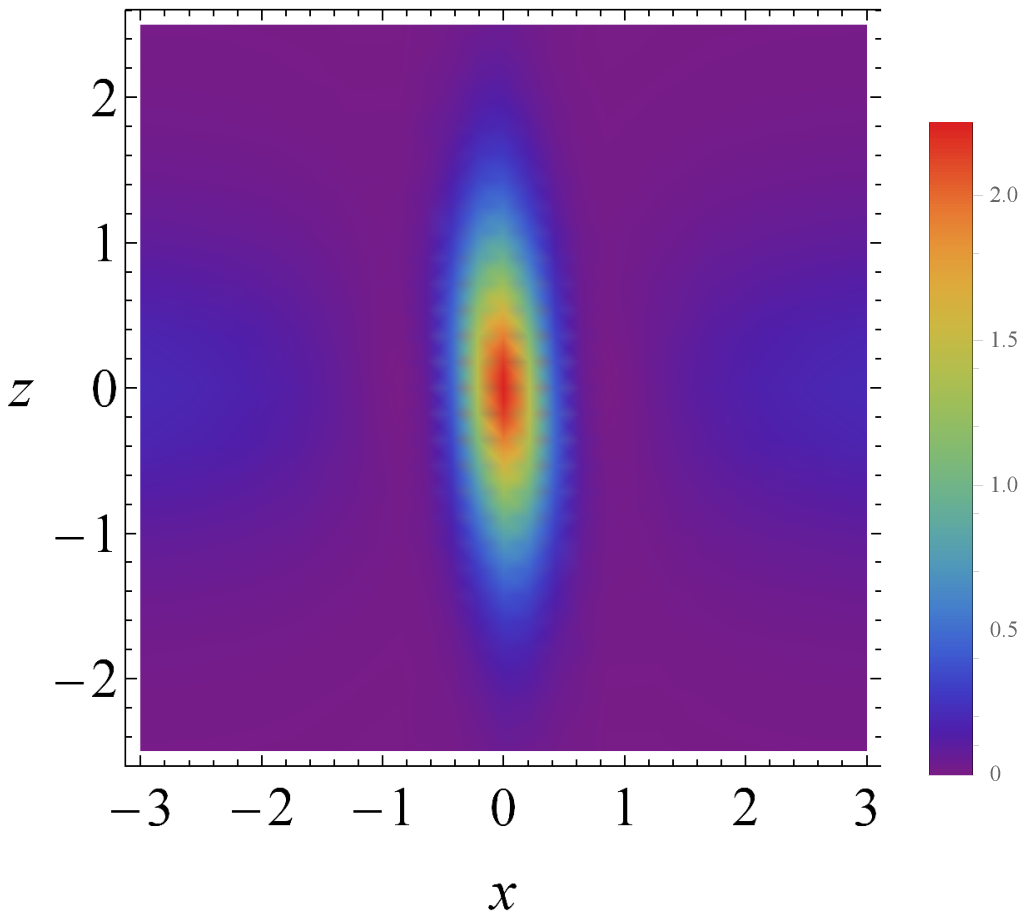}}
    \subfigure[]{\includegraphics[scale=0.57]{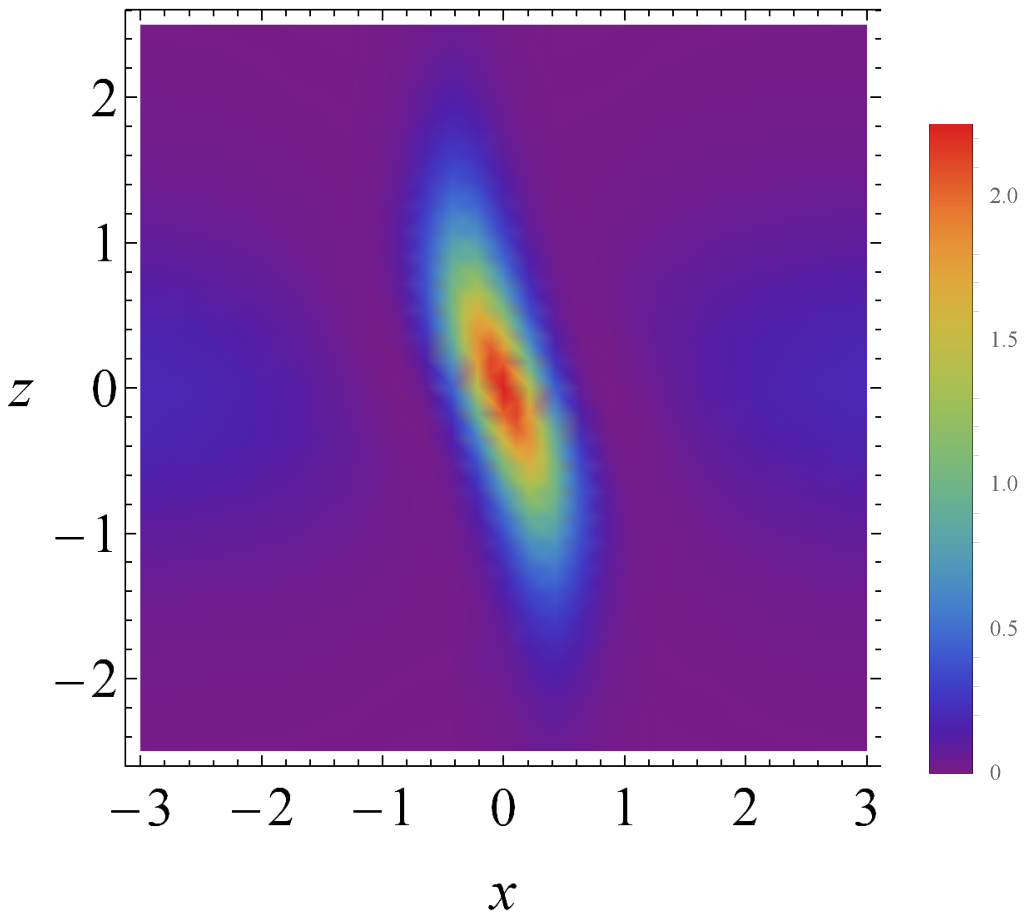}}
    \subfigure[]{\includegraphics[scale=0.57]{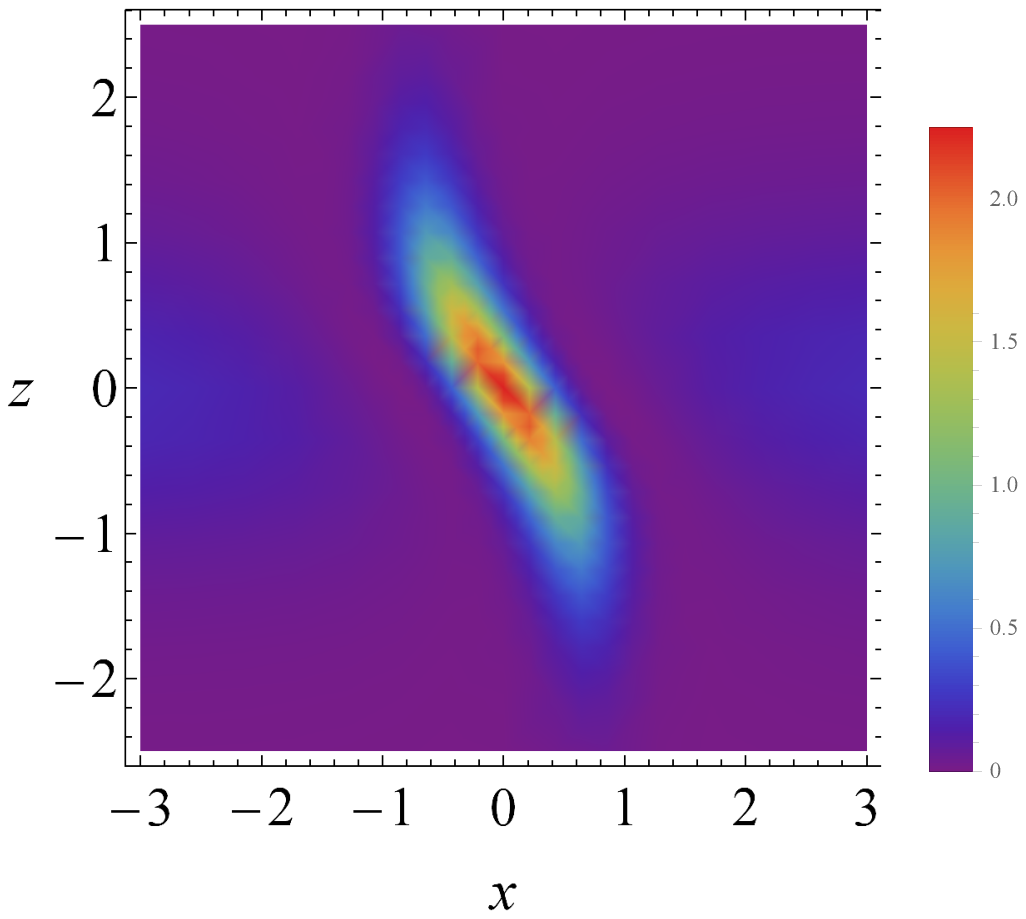}}
    \caption{\label{rogue1-int2} Contour plots for the intensity evolution of self-similar first-order rogue waves for different values of higher-order parameter, (a) $\alpha=0.1$, (b) $\alpha=0.3$ and (c) $\alpha=0.5$.}
    \end{center}
\end{figure}

The intensity expression of self-similar first- and second-order rogue waves for Eq. (\ref{vc}) reads
    \begin{align}
    \label{int3}I_{R_{1}}(x,z)&=A^2(z)\(\frac{64\xi^2+\left(-3+4\xi^2+4(6\alpha\xi+\chi)^2\right)^2}{\left(1+4\xi^2+4(6\alpha\xi+\chi)^2\right)^2}\),\\
     \label{int4}I_{R_{2}}(x,z)&=A^2(z)\(\left(\frac{F+G}{F}\right)^2+\left(\frac{\xi  H}{F}\right)^2\).
    \end{align}
The intensity profile of first- and second-order rogue waves is shown in Fig. \ref{rogue1-int1} for $\alpha=0.1,a_2 =0.8$ and $A(z)=a_0~\mbox{sech}(z)$ with $a_0=0.5$. The intensity of rogue waves is maximum at centre of the tapering region and it is localized in both directions $x$ and $z$. Further, the intensity of second-order rogue waves is more as compared to first-order rogue waves. In Fig. \ref{rogue1}(a), we present the sectional plot for the effect of tapering parameter on intensity of first-order rogue waves at centre of the tapering region. Like similaritons, the maximum intensity is more for large value of tapering amplitude, thus creating a possibility of generating high energy optical rogue waves in tapered waveguides. The rogue waves for Hirota equation has an interesting property that higher-order terms do not have any effect on the rogue wave profile at $z=0$ where the profile has its maximum except for the finite tilt with respect to the axes as we move away from $z=0$ \cite{rw3}. Here, for the first-order rogue waves, this property is preserved as shown in the Fig. \ref{rogue1}(b,c). The maximum value of intensity, at centre of the tapering region, is constant for different values of $\alpha$ (see Fig. \ref{rogue1}(b)). And, the intensity profile show little tilt in the $xz-$plane, as we move away from the central part of the tapering region as shown in the Fig. \ref{rogue1}(c). To get the better overview of intensity distribution in $xz-$plane, for different values of $\alpha$, we depicted the contour plots for the intensity profile of self-similar first-order rogue waves in Fig. \ref{rogue1-int2}. For self-similar rogue wave solutions, we have presented the effect of tapering and higher-order parameters on the intensity of first-order rogue waves only, but the same results has also been observed for second-order rogue waves.

\begin{figure}[ht!]
\begin{center}
    \subfigure[]{\includegraphics[scale=0.26]{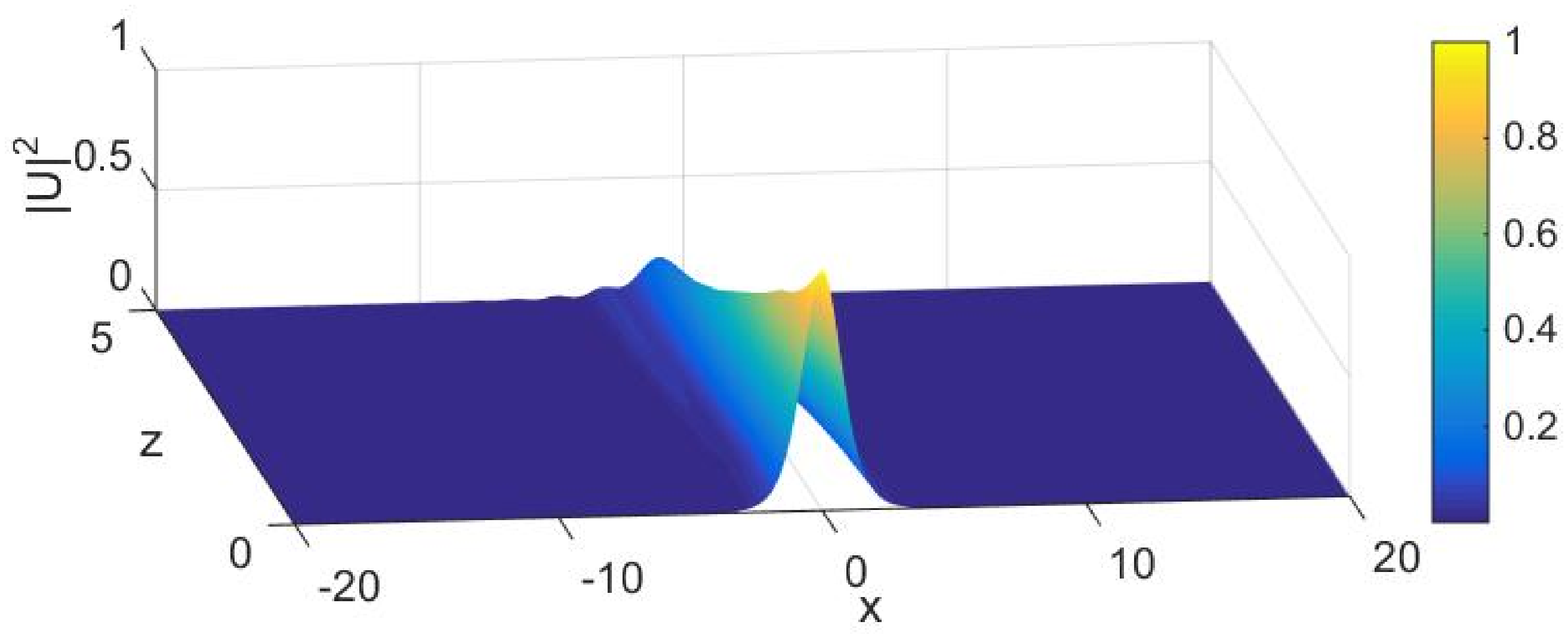}}
    \subfigure[]{\includegraphics[scale=0.26]{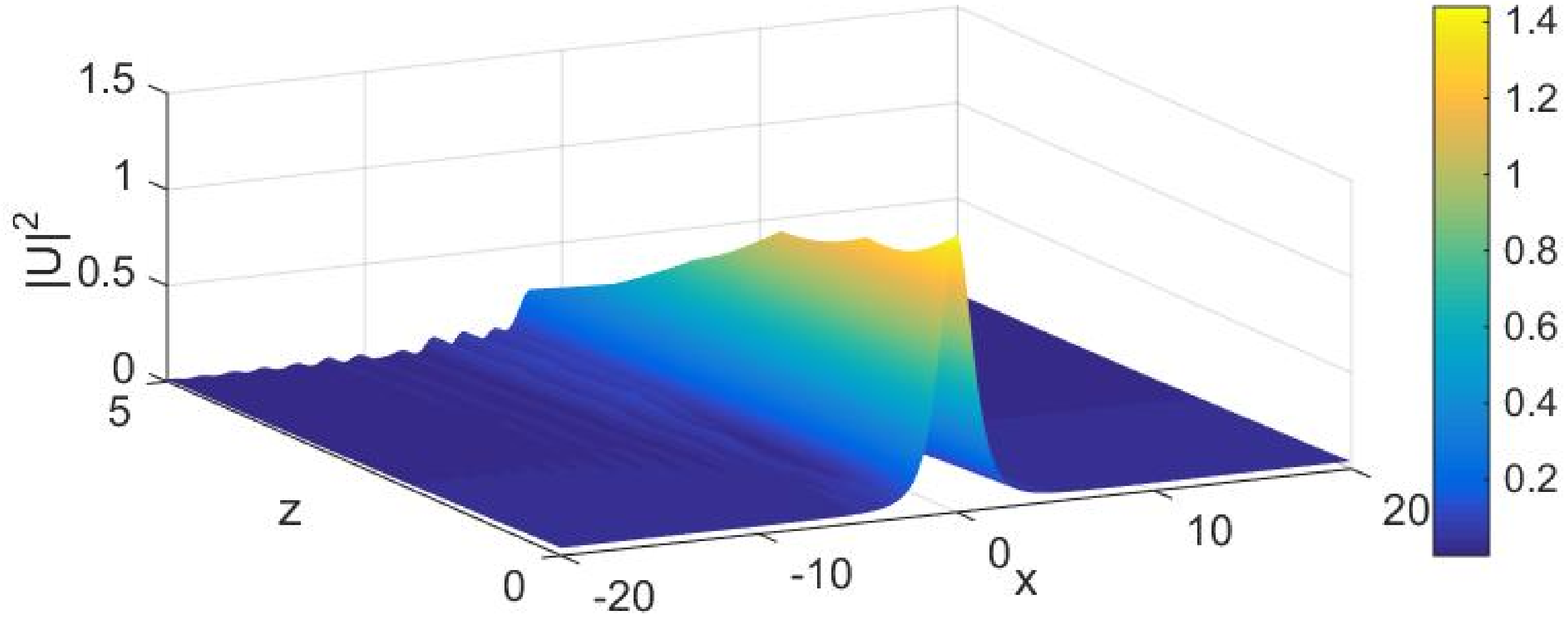}}
    \subfigure[]{\includegraphics[scale=0.27]{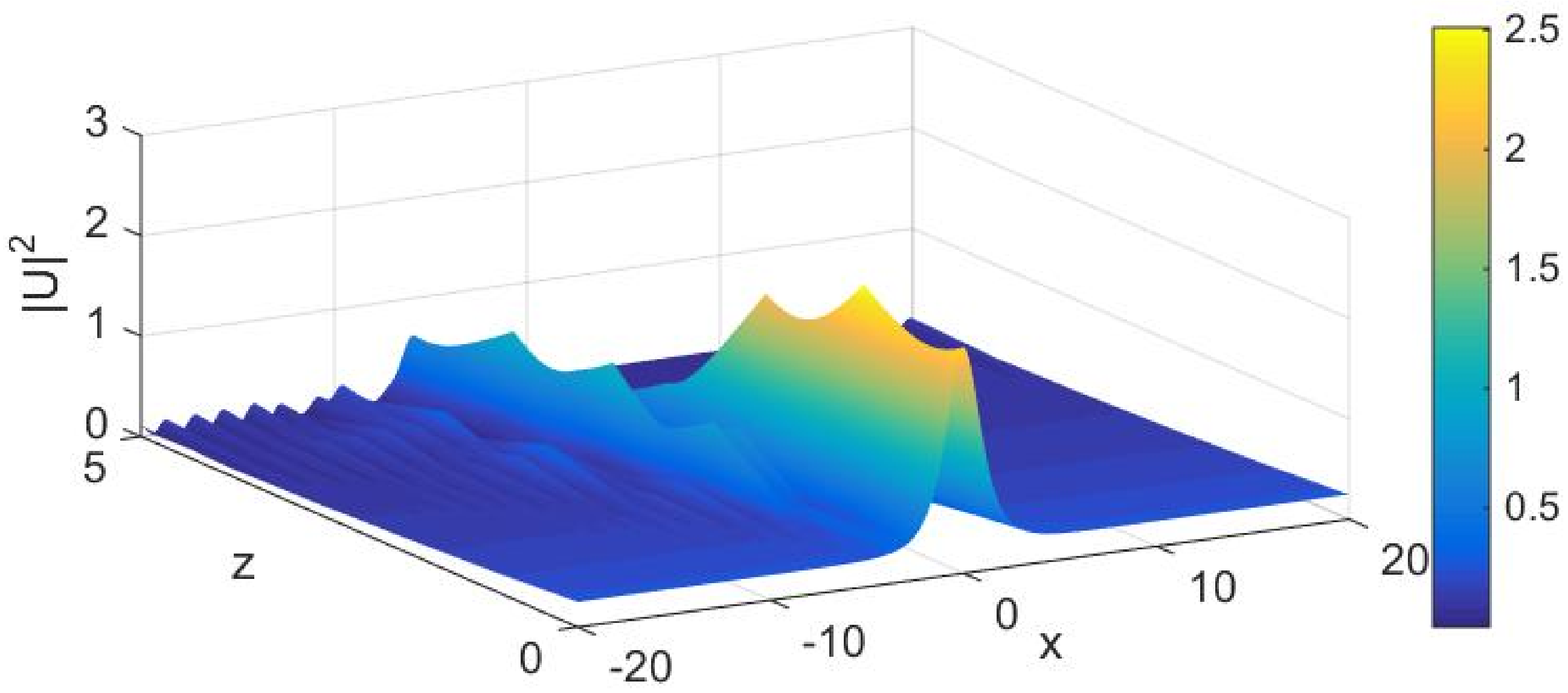}}
    \caption{\label{numer} Plot depicting the simulated intensity of soliton pulse when (a) the initial perturbation (dc-offset) $\epsilon=0$, (b) $\epsilon=0.2$, (c) $\epsilon=0.5$.}
\end{center}
\end{figure}

\section{Numerical Simulations}
Here, we have used split-step Fourier transform method to simulate the intensity of the soliton pulse for Eq. (\ref{vc}), when $A(z)=a_{0}~{\rm sech}(z)$, and we depict its evolution in Fig. \ref{numer}. The initial pulse is taken as $U(x,0)={\rm sech}(x)+\epsilon$ where $\epsilon$ signifies dc-offset. We find that the self-similar wave is stable for the dc-offset is zero. But when we feed an initial noise $\epsilon=0.2$ we notice that the self-similar wave does radiate away a small amount of particles and this radiation increases as we increase the value of dc-offset from $\epsilon=0.0$ to $\epsilon=0.8$ and if we further increase the value of perturbation, the self-similar wave becomes, highly unstable.

\section{Conclusion}
In conclusion, we have shown the possibility of producing localized optical similaritons and self-similar rogue waves for beam propagation in tapered waveguide satisfying $\mathcal{PT}$-symmetry property. It has been accomplished by first reducing the generalized HNLSE into constant coefficient HNLSE using similarity transformation and then mapping the constant coefficient HNLSE to integrable Sasa-Satsuma and Hirota equations. We manifested that the intensity of self-similar waves can be controlled efficiently through modulation in the tapered profile for judicious choice of tapering parameter. The higher-order terms imposed significant effect on the intensity of similaritons whereas, for rogue waves, maximum intensity is independent of higher-order effects except for the change in intensity distribution in $xz-$plane. Though we presented our results for sech$^2$-type tapering profile, but in general, one can study evolution of self-similar waves for any choice of tapering function. We observed same results for gaussian-type tapering function which have similar tapering profile as sech$^2$-type tapering function except to the linear profile for gain / loss function compared to the kink profile in the present case. The exact self-similar waves reported in this work may be useful for various optical processes, especially in areas such as waveguide amplifiers, nonlinear optical switches, optical couplers and ultrafast optical communications. These results may have potential applications in the experiments related to compression and amplification of ultrashort pulses \cite{tap6,tap7} and supercontinuum generation in tapered fibers \cite{tap2,tap3}.

\section{Acknowledgment}
A.G. gratefully acknowledges Science and Engineering
Research Board (SERB), Government of India for the award of SERB Start-Up Research Grant
(Young Scientists), sanction No: YSS/2015/001803, during the course of this work. H.K. and Nisha would also like to thank SERB for the award of fellowship, under the same grant, during the work tenure.

\end{document}